\documentstyle[twoside,fleqn,espcrc2]{article}%
\def\la{\langle}
\def\ra{\rangle}

\newcommand{\be}{\begin{eqnarray}}
\newcommand{\ee}{\end{eqnarray}}

\newcommand{\eq}{\begin{equation}}
\newcommand{\eqx}{\end{equation}}
\newcommand{\eqn}{\begin{eqnarray}}
\newcommand{\eqnx}{\end{eqnarray}}
\newcommand{\ben}{\begin{eqnaray}}
\newcommand{\een}{\end{eqnarray}}

\begin{document}

\title{Chiral Disorder in Two-Color QCD with Abelian External Fluxes}

\author{%
	Romuald A. Janik%
	\address{Service de Physique Th\'{e}orique,
	CEA Saclay, F-91191 Gif-Sur-Yvette, France \\[2mm]
	$^b$Marian Smoluchowski Institute  of Physics, Jagellonian
	University, 30-059 Krakow, Poland}$^{,b}$,\,%
	\refstepcounter{address}%
	 Maciej A.  Nowak$^{b,}$%
	\address{GSI, Planckstr. 1, D-64291 Darmstadt, Germany},\,%
	G\'{a}bor Papp%
	\address{CNR Department of Physics, KSU, Kent, Ohio 44242, USA \& \\
	\hspace*{1mm} HAS Research Group for Theoretical Physics,
	E\"{o}tv\"{o}s University, Budapest, Hungary}%
	,\,and
	Ismail Zahed%
	\address{Department of Physics and Astronomy, SUNY, Stony Brook, New
	York 11794, USA}
}

\begin{abstract}
\hrule
\mbox{}\\[-0.3cm]

\noindent{\bf Abstract}\\[-1mm]

We investigate the effects of several Abelian external fluxes $\phi$, on
the Euclidean Dirac spectrum of light quarks in  QCD with two colors.
Our results provide for a novel
way of assessing the pion weak decay constant directly from spectral
fluctuations, and discriminating a flux-rich from a flux-poor vacuum,
using QCD lattice simulations.\\[0.15cm]
{\em PACS}: 11.30.Rd, 12.38.Aw, 64.60.Cn\\[-0.1cm]
\hrule
\end{abstract}

\maketitle

\section{Introduction}

In two-color QCD the quarks are in the pseudoreal representation
of the flavor group~\cite{PESKIN,LEUTSMILGA}.  For two flavors,
the spontaneous breaking
of chiral symmetry is accompanied by the occurrence of five Goldstone
modes: 3 pions and a baryon and an antibaryon. The five Goldstone modes
decay weakly, with a mass that satisfies the Gell-Mann--Oakes--Renner
(GOR) relation~\cite{PESKIN}. The latter mass vanishes in the chiral limit.

In this talk we consider the effects of several Abelian external
fluxes on the light quarks in a finite Euclidean volume with two-color QCD.
We will show that these fluxes alter the quark spectral correlations near
zero virtuality in a way that is sensitive to the bulk conductance of the
disordered system~\cite{USPRL}, hence to the pion decay constant. In a finite
volume and in current lattice simulations, this provides for a novel way
to measure the pion decay constant. Our analysis in QCD
will parallel recent analysis of electrons in disordered
metals~\cite{MONTAM}. A more thorough discussion can be found in~\cite{USXX}.

\section{QCD with Abelian fluxes}

In a finite Euclidean volume $V=L^4$ pierced by Abelian fluxes
$\phi_{\mu}=(\phi_1,\phi_2,\phi_3,\phi_4)$,
the light quarks in the background gluon field $A$ satisfy the following
Dirac equation
\be
i\nabla \!\!\!\!/ [A] \,q_k =\lambda_k [A, \phi] \, q_k \,\,.
\label{01}
\ee
subject to the boundary condition
\be
q_k (x+L_{\mu}) =-e^{i2\pi \phi_{\mu}}\,\,q_k (x) \,.
\label{02}
\ee
The $\phi$'s are given in units of a
flux quantum $h/e_q$ set to 1 for convenience.
Through (\ref{02}) the quark spectrum (\ref{01}) depends explicitly on $\phi$.

The probability $p(t, \phi)$ for a light quark to start at $x(0)$ in $V$ and
return  back to the same position $x(t)$ after a proper time duration $t$, is
\be
p(t, \phi )= \frac {V^2}N
\Big\la |\la x(0)|e^{i(i\nabla \!\!\!\!/[A] +im)|t|}|x(0)\ra|^2\Big\ra_A.
\label{1}
\ee
The averaging in (\ref{1}) is over all gluon
configurations using the unquenched two-color QCD measure
%. The normalization in (\ref{1}).
The normalization
is per state, where $N$ is the total number
of quark states in the four-volume $V$. Eq.~(\ref{1}) may be
written in terms of the standard Euclidean propagators for the quark field,
\be
&&p(t, \phi ) = \frac {V^2}N \lim_{y\to x}
\int \frac {d\lambda_1d\lambda_2}{(2\pi)^2}
\,e^{-i(\lambda_1-\lambda_2) |t|}\nonumber\\&&
\times\Big\la {\rm Tr}\left( S(x,y;z_1,\phi)
S^{\dagger} (x,y; z_2,\phi)\right)\Big\ra_A
\label{des2}
\ee
generalizing to $\phi \neq 0$ the results in~\cite{USPRL}.
Here  $z_{1,2}=m-i\lambda_{1,2}$, and
\be
S(x+L_{\mu}, y;z,\phi) =-e^{i2\pi\phi_{\mu}}\,\,
S(x,y;z,\phi) \,.
\label{bound}
\ee

The correlation function in (\ref{des2}) relates (in general) to
an analytically continued pseudoscalar correlation function, as
the eigenvalues $q_k$ and $-\gamma_5 q_k$ are
pair-degenerate (chiral)~\cite{USPRL}.
For two-color QCD there is an extra symmetry~\cite{PESKIN,LEUTSMILGA}
that makes the pseudoscalar correlation function degenerate with certain
diquark correlation functions in the flux-free case. Indeed, for $\phi=0$
and ${\bf K}=-\tau^2C{K}$ the eigenvalues $q_k$ and
${\bf K}\, q_k$  are  also
pair-degenerate. Here $C$ is the charge-conjugation matrix,
$\tau^2$ the color matrix and $K$ the (right-left) complex-conjugation.

For $\phi\neq 0$, the first symmetry (chiral) is retained while the second
one is upset~\cite{NOTEX}. The quarks are now required to be in the complex
representation, except for the case where $\phi$ is half-integer. The
fluxes add in the cooperon channel (diquark)
with net Abelian charge 2, but cancel in the diffuson (pseudoscalar)
with net Abelian charge 0. In the semiclassical limit, both contribute
to (\ref{des2}) as discussed in~\cite{USXX}.
Using the GOR relation $F^2m_{G}^2=m\Sigma$
and the analytical continuation $m\rightarrow m-i\lambda/2$,
the expectation value ${\bf C}_G\equiv\la{\rm Tr} SS^\dagger\ra$, in
(\ref{des2}) simplifies to
%\vspace*{-1mm}
\be
&&{\bf C}_{G} (x,y;z, \phi) \approx +\frac 1{2V} \sum_{n_\mu}
e^{iQ\cdot (x\!-\!y)}
\nonumber\\[-1mm]
&&\hspace*{-5mm}\times\left(\frac {2\Sigma}{-i\lambda\!+\!2m\!+\!
%DQ(0)^2} +
DQ^2} +
\frac {2\Sigma}{-i\lambda\!+\!2m\!+\!
%D{Q(\phi)}^2}
D\tilde{Q}^2}
\right)
\label{des6}
\ee
with the diffusion constant $D=2F^2/\Sigma$,
and $Q_{\mu} = n_{\mu}2\pi/L$ and
%$(n_{\mu}\!+\!2\phi)2\pi/L$ and
$\tilde{Q}_{\mu}=(n_{\mu}\!+\!2\phi_{\mu}) 2\pi/L$.
%where $n_{\mu}$ are integers.
$\Sigma=|\la \overline{q} q\ra |$ and $F$ is the weak decay
constant for the Goldstone modes of mass $m_G$~\cite{NOTEXX}.

Since $E/\Delta=N$ and $\rho=1/\Delta V$,
with $\Sigma=\pi \rho$, we conclude after
a contour integration that
\be
p(t, \phi) = \frac 12 e^{-2m|t|}\sum_{n_\mu} \left( e^{-
%DQ(0)^2
DQ^2
|t|} +
e^{-
%D{Q(\phi)}^2
D\tilde{Q}^2
|t|}\right)\,.
\label{des7}
\ee
The cooperon contribution is periodic in the flux $\phi$
with periodicity $\phi=0, \pm 1/2, \pm 1, ...$.
The spectrum does not discriminate
between bosonic or fermionic boundary conditions in the flux-free
case. The cooperon contribution $p_C(t,\phi)$, (second term in {\ref{des7})
may be rewritten using Poisson's
resummation formula as
\be
%p_{C} (t,\phi)
p_C\!=\frac V{2(4\pi D t)^2} %\nonumber \\
%&&\hspace*{3mm}\times
\sum_{[l]}\!e^{-2m|t|-\frac{l_{\mu}^2 L^2}{4D|t|}}
	\,\cos{(4\pi l_{\mu} \phi_{\mu})}
\label{pois}
\ee
with integer $l$'s. This result is in agreement with the one derived
by Montambaux~\cite{MONTAM} in disordered metals in lower dimensions.
The flux-accumulation in the cooperon part implies
changes in the spectral correlations of the light quarks
as we now show.

\section{Spectral Rigidity}

To describe the spectral correlations associated with (\ref{01})
we will use semi-classical arguments for the two-point correlation
function $R(s, \phi)$ of the density of
eigenvalues~\cite{USPRL,MONTAM,IMRY}. Its spectral
form factor $K(t,\phi )$ is defined as
\be
R(s , \phi) =\int_{-\infty}^{+\infty} dt\, e^{is\Delta t}\, K(t, \phi)
\label{spectral2}
\ee
where $\Delta=1/\rho V$ is a typical quantum spacing at zero virtuality
and in the absence of a flux. For diffusive quarks in two-color QCD,
$K(t,\phi )$ relates to the return probability
through~\cite{USPRL,MONTAM,IMRY},
\be
K(t, \phi) \approx \frac{\Delta^2 |t|}{(2\pi )^2}\, p(t, \phi)\,\,.
\ee

For simplicity, consider the
case $\phi_{\mu}=(0,0,0,\phi)$ with only one-flux retained. For long proper
times, the zero modes along the $1,2,3$ directions contribute only, giving
\be
&&p(t, \phi) =\frac 12 e^{-2m |t|}\,\times \nonumber\\
&&\hspace{-7mm}\sum_n\left( e^{-4\pi^2 n^2 \sigma_L |t/t_H| }+\!
e^{-4\pi^2 (n+2\phi)^2\sigma_L |t/t_H|}\right)
\label{per}
\ee
which is the analogue of a diffusion in d=1.
$\sigma_L=D/(\Delta L^2)=2(FL)^2/\pi $
is the dimensionless conductance, where $F^2$ can be regarded as
the conductivity characteristic of the
flow of the isoaxial-charge in 4+1-dimensions~\cite{USPRL}.
For $t<\tau_{\rm erg}=L^2/D$ the
diffusive paths are short and do not accumulate enough flux. The spectral
rigidity defined as~\cite{USPRL,USXX} (and references therein)
\be
\Sigma_2(N,\phi) =\int_{-N}^{+N}\, ds\,(N-|s|)\,R(s, \phi)
\ee
is in our case
\be
\Sigma_2 (N, \phi) =
\frac 1{2\pi^2}\!\sum_n {\rm ln}\!\left[\left( 1\!+\!
\frac {N^2}{\alpha_n^2}\right)
\left( 1\!+\!\frac {N^2}{\tilde{\alpha}_n^2}\right)\right]
\label{varx}
\ee
with $\alpha_n=\alpha+4\pi^2\sigma_L n^2$ and
$\tilde{\alpha}_n=\alpha+4\pi^2\sigma_L (n+2\phi)^2$.
For $N,\sigma_L\gg \alpha$, (\ref{varx}) simplifies to
\be
\Sigma_2 (N, \phi)\!=\!\Sigma_2 (N, 0)\!-\!\frac 1{\pi^2}{\rm ln}\!
\left(1\!+\!4\frac {\sigma_L}{\alpha}\, {\rm sin}^2 2\pi \phi \right)
\label{large}
\ee
in agreement with a result derived by Montambaux~\cite{MONTAM}
in the context of disordered metals.
The conductance $\sigma_L$, of the chiral vacuum is
directly accessible from the spectral rigidity through (\ref{large})
providing for a direct measurement of this important quantity
in disordered QCD.
When combined with the measurement of the chiral condensate $\Sigma$,
this observation allows for a novel determination
of the pion weak-decay constant $F$ solely from investigations
of the quark spectrum using lattice QCD.

\section{QCD Vacuum}

Since the quark return probability and the spectral rigidity are
sensitive to flux-variations in a finite Euclidean volume, they
could be  used to probe the flux-content of the two-color QCD vacuum,
provided that the fluxes upset the pseudoreal character of the quark
fields. Indeed, if we consider an Abelian flux-disordered vacuum along
the lines we have so far discussed, and characterized by a Gaussian
distributed flux with a mean
\be
\la\la\phi_\mu\phi_\nu\ra\ra=\kappa_\mu^2\delta_{\mu\nu}
\label{measure}
\ee
then the quark return
probability can be easily estimated from (\ref{des7}) using the Poisson form
(\ref{pois}) for the cooperon part. If we split the quark return
probability $p =p_D+p_C$, then the diffuson part $p_D$, is flux-insensitive
\be
\la\la p_D (t, \phi)\ra\ra =\frac V{2(4\pi D t)^2}\,\,e^{-2m|t|}
\label{vac1}
\ee
while the cooperon part $p_C(t,\phi)$, is flux-sensitive
\be
\la\la p_{C}\ra\ra\!=\frac V{2(4\pi D t)^2}
\sum_{[l]}\!e^{-2m|t|-\frac{l_{\mu}^2 L^2}{4D|t|}
	-8\pi^2 \kappa_{\mu}^2 l_{\mu}^2}.
\label{vac2}
\ee
We note that the pe\-rio\-dic\-ity in
$\phi=$$0, \pm 1/2, \pm 1, ...$ of the quark return probability
implies that the latter is likely to be insensitive to a $Z_2$
flux-disordered vacuum. If these effects extend to an Abelian flux-rich
vacuum, a simple way to detect them is to measure the relative ratio of the
quark return probabilities for $N_c=2$ and $N_c=3$. A flux sensitivity
implies a t-dependent ratio close to $1/2$ (as opposed to 1), assuming that
the vacuum structure does not change appreciably from two to three colors.

\section*{\bf Acknowledgments}

This work was supported in part by the US DOE grants DE-FG-88ER40388 and
DE-FG02-86ER40251, by the Polish Government Project (KBN) grants
2P03B01917
%2P03B04412
and 2P03B00814 and by the Hungarian grant OTKA-T022931.


\begin{thebibliography}{99}
\bibitem{NOTEX}
Throughout we assume that the light baryon-antibaryon
fluctuations do not dynamically `screen' the Abelian
fluxes. This is evident at $\phi= 1/2$ where there is
only a net change in the statistics. This
assumption is not needed in the quenched case.
\bibitem{NOTEXX}
The implicit flux dependence of the present parameters at next
to leading order is subsumed.
\end{thebibliography}
\end{document}